\documentstyle[preprint,pre,aps,eqsecnum]{revtex}

\begin{document}
\draft


\title{First Order Phase Transition in Intermediate Energy Heavy Ion
Collisions}

\author{Jicai Pan, Subal Das Gupta, and Martin Grant}

\address{
Physics Department, McGill University,
3600 University St., Montr{\'e}al, Qu{\'e}bec \\ Canada H3A 2T8}

\date{ \today } 

\maketitle

\begin{abstract}
We model the disassembly of an excited nuclear system formed as a
result of a heavy ion collision.  We find that, as the beam energy in
central collisions in varied, the dissociating system crosses a
liquid-gas coexistence curve, resulting in a first-order phase
transition.  Accessible experimental signatures are identified: a peak
in specific heat, a power-law yield for composites, and a maximum in
the second moment of the yield distribution.  
\end{abstract}





Nuclear matter is a fictitious arbitrarily large $N=Z$ system in which
the Coulomb interaction is switched off.  Mean-field theory for nuclear
matter has been done many times, and it is well-known that such a
system shows a Van-der-Waals-type liquid-gas phase transition.  It was
suggested in the early eighties \cite{Curtin83,Finn82} that in heavy
ion collisions at intermediate energies one might be able to probe this
liquid-gas phase transition region.  In heavy ion collisions, matter
would be heated as well as compressed.  This compressed blob would then
expand passing through the liquid-gas coexistence phase.  One might be
able to extract information about this region from selected
experimental data.  Earlier, the Purdue group had conjectured that the
break up of large nuclei by energetic protons would show signatures of
critical phenomena \cite{Finn82}.

There are several complications which make the study of phase
transitions in nuclei difficult.  The systems are small, thus
singularities get replaced by broad peaks.  The collision is over
quickly.  The existence of thermal equilibrium has sometimes been
questioned and replaced by quite complicated transport equation
approaches.  Often one has been content to do calculations where the
objective is to fit the experimental data.  Once this is achieved the
question of a possible liquid-gas phase transition is not addressed.
There are models where such questions are irrelevant, or at least very
indirect, such as various models based on sequential decays.  The
literature on calculations for fragment yields in intermediate energy
heavy ion collisions is huge.  We will not attempt to mention all
approaches.

In this paper we will focus on the liquid-gas phase transition using a
lattice gas model \cite{Pan95}.  Previously we have used the model to
fit data on central collisions \cite{Pan95}, on peripheral collisions
\cite{Beaulieu96}, and for central Au on Au collisions
\cite{DasGupta96}.  The model gives a fair description of data in those
instances --- does it say anything definite about the experimental
signature of liquid-gas phase transitions?

In the lattice gas model we place $n$ nucleons in $N$ cubes, where $n$
is the number of nucleons in the disassembling system, and
$N/n=\rho_0/\rho_f$, where $\rho_0$ is the normal nuclear density, and
the disassembly is to be calculated at $\rho_f$, the ``freeze-out''
density beyond which nucleons are too far apart to interact.  An
attractive nearest-neighbor interaction is assumed.  We place the $n$
nucleons in $N$ cubes by Monte-Carlo sampling using the Metropolis
algorithm.  Once the nucleons have been placed we also ascribe to each
of them a momentum.  The momenta are generated by a Monte-Carlo
sampling of a Maxwell-Boltzmann distribution.  Various observables can
be calculated in a straightforward fashion.  Experiments usually
measure the distributions of cluster sizes, i.e., the yield $Y(Z)$,
which is a function of the number of protons in the composite $Z$.  In
our model, two neighboring nucleons are considered to be part of the
same cluster if their relative kinetic energy is insufficient to
overcome the attractive bond:  $p_r^2/2\mu+\epsilon<0.$  This can be
immediately turned into a temperature and $\epsilon$ dependent bonding
probability \cite{Pan95,DasGupta96} much like Coniglio-Klein's
\cite{Coniglio80}.  The prescription allows us to calculate the
distribution of clusters.

The equation of state in the lattice gas model has been well studied in
condensed matter physics.  The grand canonical ensemble for the lattice
gas corresponds to a three-dimensional Ising model in the presence of a
magnetic field.  It is thereby possible to translate many well known
results to the present situation.  Fig.~1 depicts the phase diagram of
the lattice-gas model in the thermodynamic limit,
which is adapted from Ref.~\cite{Kertesz83}.  
In drawing the coexistence curve DCE, the series expansion given in
Ref.~\cite{domb} was used.
The point ``C'' is the thermal critical point which occurs at $T \approx
1.1275|\epsilon|$ and density $\rho_f/\rho_0 \equiv 1/2$.  The
coexistence curve is DCE, and CB is the line along which percolation
sets in.  The line CB is only slightly different in our and
Coniglio-Klein's prescription.

In an experimental situation involving the collision of heavy ions, one
gates on central collisions and varies the beam energy.  At low beam
energy, the freeze-out density will be below the co-existence curve.
As the beam energy increases, this point will cross the coexistence
curve (indicated by an arrow in Fig.~1).  Note that there is no reason
for the system, under heating, to ``tune itself'' to the second-order
point at density $\rho = 1/2$, since there is no symmetry between the
clustered and unclustered phases involved in the heavy-ion collision.
Indeed, in statistical models of disassembly of which we know
\cite{Bondorf95,Gross97}, the freeze-out density is less than
$0.5\rho_0$.  In our model, we find that the data are best fitted by a
$\rho_f$ between $0.3\rho_0$ and $0.4\rho_0$.  Therefore the phase
transition is {\it first order\/}.

As one crosses the coexistence curve, experiments would see various
signatures of the first-order transition.  In the thermodynamic limit
these are well defined.  For example, since first-derivatives of the
free energy are discontinuous in the thermodynamic limit, there are
delta-function peaks in second derivatives 
\cite{Pippard+Callen,Note}, 
e.g., in the heat capacity
at constant 
pressure $C_p$, 
and in the isothermal compressibility $\kappa_T$.
There is also a clear signature of the transition in the
density-density fluctuation correlation function $\Gamma(r)$ (whose
integral gives the compressibility by a thermodynamic sum rule
\cite{Goldenfeld92} $\int d\vec r \; \Gamma(r) \propto \kappa_T$).
This correlation function decays exponentially above and below the
transition point.  At the transition point itself, however, $\Gamma(r)$
is flat 
\cite{Goldenfeld92,Privman83}, 
as is evident from, for example, the sum rule.

This singular behavior is smeared out in finite-size systems
\cite{Privman83}.  Instead of delta-function singularities, 
$\kappa_T$ and $C_p$
have broad Gaussian peaks at the first order
transition, whose height is proportional to the system's volume $\sim
L^3$.  Indeed, these features are somewhat analogous to those at a
continuous second-order transition: In the thermodynamic limit 
at a continuous transition, $C_p$ and $\kappa_T$ have power-law 
singularities, 
but in a finite-size system, these singularities are replaced by
bumps of height $\sim L^{\gamma/\nu}$,
where $\gamma$ and $\nu$ are critical exponents.  This
similarity implies why it is often difficult to distinguish the two
types of transitions in a finite-size system
\cite{Goldenfeld92,Privman83}.  

While $C_p$ and $\kappa_T$ are both infinite at a first-order
liquid-gas transition, the heat capacity at constant volume $C_v$ has
only a step discontinuity in the thermodynamic limit, giving a bump in
a finite-size system.  The extent of the discontinuity is given in
terms of thermodynamic relations in Ref.~\cite{Heller67}.  This is in
contrast to the more pronounced behavior at a second-order transition
where $C_v \sim L^{\alpha/\nu}$, and $\alpha$ is a critical exponent.

In Fig.~2 we show results of our numerical calculation.  We fix $N$
at $7^3$ and vary $n$ to obtain a variable freeze-out density.  Half
of the nucleons are labeled as protons.  Although this system is
small, there is a well defined bump in $C_v$ signifying the
transition.  An arrow marks the nearby point at which the transition
occurs in the thermodynamic limit.

In nuclear experiments locating the peak in $C_v$ as the beam energy
increases is difficult, although in recent years tremendous progress
has been made in the measurement of temperature \cite{Huang97}.
However, other characteristics of the transition appear which are
more readily measurable.  In particular, the yield $Y(Z)$ is readily
measurable, which gives the distribution of clusters of charge $Z$,
$n(Z)$.  The probability of clusters of a given size is related
(through its second moment), to the density correlation function
\cite{Kertesz83,Stauffer92,Huse87}:  
Roughly, if the clusters are not fractals,
$Z^2 n(Z) \sim \Gamma (r)$, in the disordered phase, where $r$ is
the diameter of a cluster of charge $Z$.  At the transition, a
droplet of thermodynamic size spans the system.  Hence the yield
includes an infinite cluster, or in percolation language, a cluster
which spans the system.  The remainder of the distribution describes
fluctuations in that phase, giving the density correlation function,
whose integral is the compressibility.  The delta-function peak in
the compressibility implies that the correlation function, and hence
the yield, is broad.  Following standard practice \cite{Pratt95},
the yield is fit to a power law form $Y(Z) \propto 1/Z^{\tau}$,
giving an effective exponent $\tau$, even when the distribution has
deviated from a power law.  In fact, at a {\it continuous\/} phase
transition, $\tau = 2 + \beta/(\beta + \gamma)$, where $\beta$ is a
critical exponent.  Here, however, where the transition is
first-order, the correlation function decays exponentially in either
bulk phase (i.e., the effective  $\tau \rightarrow \infty$ above or
below the transition), while, since the correlation function is flat
at the transition itself, and is related to the second moment of the
cluster distribution, $\tau = 2$ there, for an infinite system.  Of
course, in a finite-size system, there will be an effective $\tau$,
which is neither two nor infinity, so that the transition will look
somewhat analogous to a continuous transition.

It is also useful to consider the second moment of the cluster
distribution function, $S_2=\sum'_A A^2n(A)/n$, where $n(A)$ is the
number of nucleons with mass number $A$, and $n$ is the total number
of nucleons.  The primed sum excludes the largest cluster, and so we
expect it \cite{Huse87} to be proportional to the compressibility
$\kappa_T$.
The usefulness of the second moment was emphasized by Campi
\cite{Campi88}.  

Fig.~2 shows that these quantities do indeed provide clear
experimental signatures of the first-order transition in a
finite-size system.  The crossing of the coexistence line is evident
in the minimum value of the effective $\tau$, which is close to 2,
and the prominent maximum in $S_2$ and $C_v$.  All of these occur at
approximately the same point, close to the transition in the
thermodynamic limit, indicated by an arrow.

For our analysis, it is important that the freeze-out density is on
the low density side of the coexistence curve.  Beyond the critical
point, ``C'', the 
bump 
in $C_v$ continues on the line CE whereas
the minimum of $\tau$ will follow line CB, which corresponds to a
change in short-range order.  We should also note that very large
Coulomb forces will alter this picture.  We now describe briefly
those effects, following Ref.~\cite{DasGupta96}.  In our lattice gas
model one generates an event in which the nucleons have been placed
with appropriate momenta.  Using our prescription we can immediately
obtain the cluster distribution.  Starting from this configuration
we can propagate the system by molecular dynamics with a suitably
chosen short range nucleon-nucleon interaction (to correspond to the
nearest neighbor interaction) with or without the Coulomb
interaction.  By propagating molecular dynamics for a considerable
time one can unambiguously obtain the cluster distribution, and
determine the effect of the Coulomb interaction.
Ref.~\cite{DasGupta96} concluded that without the Coulomb
interaction there is little difference between lattice gas results
and molecular dynamics results, but that Coulomb interactions can be
important for large mass numbers.  For mass number $A=85$, the
Coulomb interaction is a small perturbation, while at $A=394$ it is
so strong as to totally alter the picture.  In the latter case there
is no minimum in $\tau$; at a temperature of 1 MeV, the value of
$\tau$ is slightly higher than 1 and it continues to rise
monotonically with temperature.  We have now studied this in much
greater detail and find that a minimum in the value of $\tau$
continues to be obtained for mass number as large as $A=200$.
Details of this as well as further applications of the lattice gas
model will be published in a longer paper.

To conclude, we have modeled the disassembly of nuclear matter
following a heavy ion collision.  We find that the transition is
first order, with the standard signatures of such a phase
transition.

We thank Charles Gale, Martin Zuckermann, and Hong Guo for useful
discussions.  This work is supported in part by the Natural Sciences
and Engineering Research Council of Canada and by {\it le Fonds pour
la Formation de Chercheurs et l'Aide \`a la Recherche du
Qu\'ebec\/}.

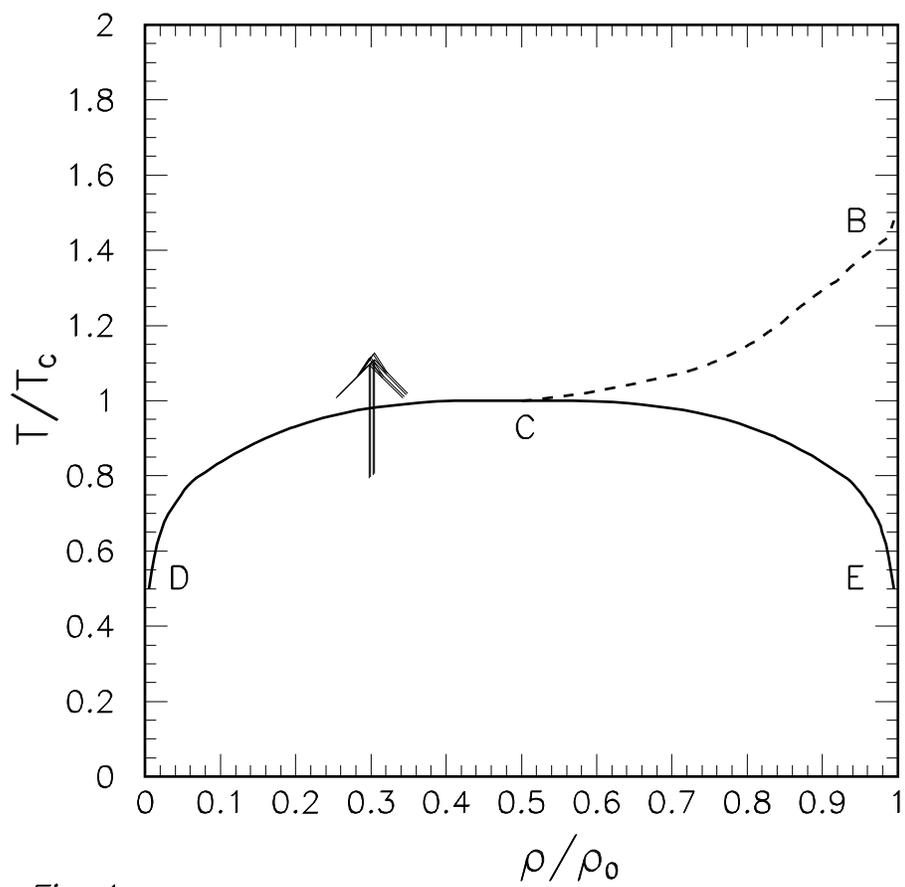
\begin{figure} 
\caption{ 
Phase diagram of three dimensional lattice gas
model.  The line DCE is the coexistence curve.  CB is the percolation
line.  The arrow demonstrates the crossing of the coexistence line as
the beam energy for central collisions increases.  In the Ising model
DCE is the line of spontaneous magnetization.  With the usual
convention the point 0 on the abscissa corresponds to magnetization
$-M$ and the point 1 to $M$.  
} 
\label{phase_diagram} 
\end{figure}

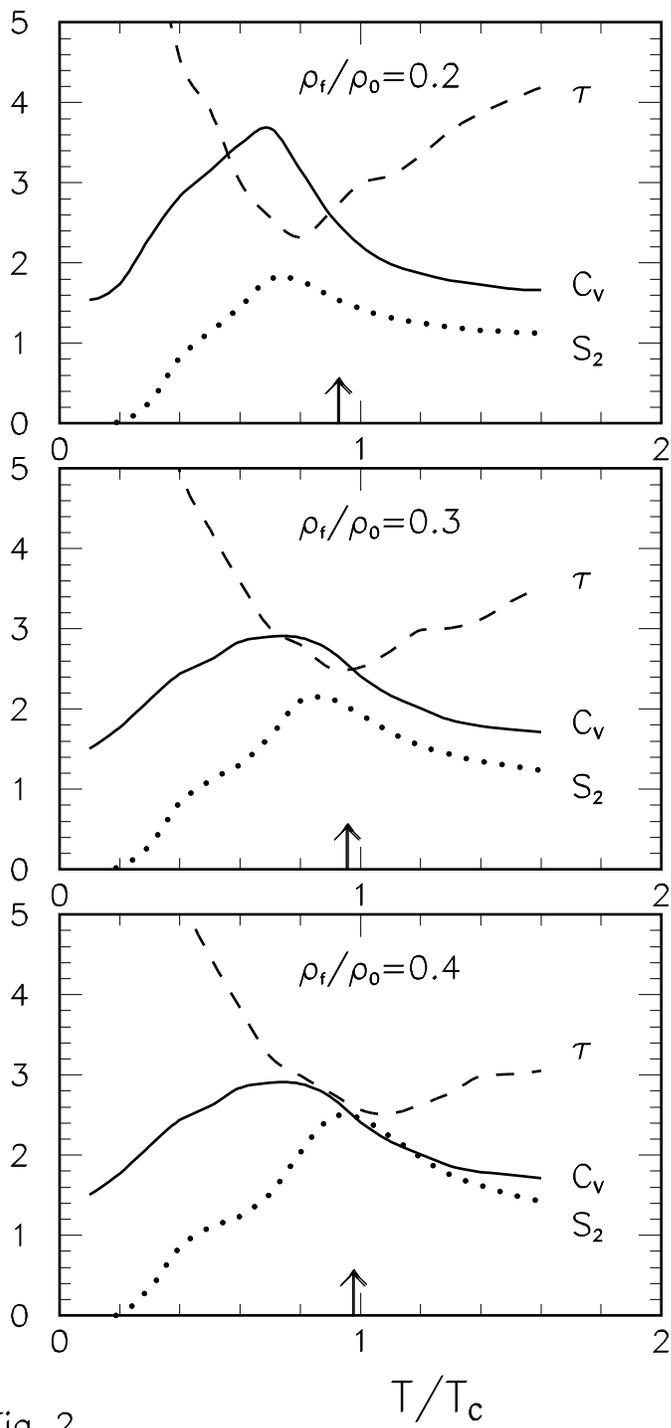
\begin{figure} 
\caption{
Curves for $\tau$, $C_v$, and $S_2$ at
$\rho_f/\rho_0 = 0.2$, $0.3$, and $0.4$.  At each point 1000 events
were taken.  $C_v$ is in units of $k_{\rm B}$; the kinetic energy
contributes 1.5 to it at all temperatures; $T_c \approx 1.1275
|\epsilon|$.  
} 
\label{heat_capacities} 
\end{figure}

\end{document}